%% file: main.tex
\pdfoutput=1
\documentclass{article}
\usepackage{spconf,amsmath,graphicx}

\input{macros.tex}

\input{tikz_style.tex}
\input{symbols.tex}

\input{glossaries}

\title{On Word Error Rate Definitions and their Efficient Computation for 
Multi-Speaker Speech Recognition Systems}
\newcommand{\upb}{$^1$}
\newcommand{\ntt}{$^2$}
\name{\tab{Thilo von Neumann\upb, Christoph Boeddeker\upb, Keisuke Kinoshita\ntt, Marc Delcroix\ntt, \\Reinhold Haeb-Umbach\upb\vspace{-1em}}}
\address{\upb Paderborn University\quad \ntt NTT corporation, Japan}
\begin{document}
\ninept
\maketitle
\begin{abstract}
	We propose a general framework to compute the word error rate (WER) of ASR systems that process recordings containing multiple speakers at their input and that produce multiple output word sequences (MIMO).
	Such ASR systems are typically required, e.g., for meeting transcription.
	We provide an efficient implementation based on a dynamic programming search in a multi-dimensional Levenshtein distance tensor under the constraint that a reference utterance must be matched consistently with one hypothesis output. 
	This also results in an efficient implementation of the ORC WER which previously suffered from exponential complexity.
	We give an overview of commonly used WER definitions for multi-speaker scenarios and show that they are specializations of the above MIMO WER tuned to particular application scenarios. 
	We conclude with a  discussion of the pros and cons of the various WER definitions and a recommendation when to use which.
\end{abstract}
\begin{keywords}
word error rate, meeting recognition, Levenshtein distance
\end{keywords}
\section{Introduction}
\label{sec:intro}

While \gls{ASR} systems may be complex, at least their generally accepted performance measure, the \gls{WER}, is seemingly simple. It is given by the Levenshtein distance between the recognized word sequence and the ground truth transcription divided by the number of words in the ground truth transcription.
The Levenshtein 
distance is defined as the minimal number of substitution, insertion and deletion operations required to turn one word string into another. The distance can be efficiently computed by use of dynamic programming, and tools, such as NIST's \verb!sclite! \cite{Fiscus_NISTScoringToolkit}, are widely used in the community.  

Conventional \gls{ASR} systems process recordings of a single speaker at the input and output a single transcription, hence we call them SISO.
However, today's \gls{ASR} systems have emerged from SISO systems to systems that process recordings of multiple speakers at their input (potentially overlapping), and produce multiple output word sequences, e.g., for meeting transcription.
We call these systems MIMO.

The final goal for a MIMO transcription system is answering the question who spoke what and when, i.e., providing both temporal information and speaker labels together with the transcribed words \cite{Kanda2020_JointSpeakerCounting,Raj2021_IntegrationSpeechSeparation}.
The input to a MIMO system can contain speech overlaps, which poses a challenge not only for ASR but also for WER computation, where the conventional WER definitions are not applicable.
Many works on ASR, however, focus only on sub-problems and provide varying degrees of detail, e.g., neglecting speaker identification \cite{Yoshioka2018_RecognizingOverlappedSpeech,Chen2020_ContinuousSpeechSeparation,Kanda2020_SerializedOutputTraining,Sklyar2022_MultiTurnRNNTStreaming}, temporal information \cite{Kanda2020_SerializedOutputTraining,Kanda2022_StreamingMultiTalkerASR} or even utterance order \cite{Kanda2020_SerializedOutputTraining}.
Evaluating such systems poses an additional challenge as there is no clearcut definition of WER that serves all purposes.

Consequently, a number of WER definitions have emerged that are tailored to specific use cases, such as \gls{uWER} \cite{Chen2020_ContinuousSpeechSeparation}, \gls{SA-WER} \cite{nist_2009_2009}, \gls{cpWER} \cite{Watanabe2020_CHiME6ChallengeTackling}, and \gls{ORC WER} \cite{Sklyar2022_MultiTurnRNNTStreaming}.
Which of these \gls{WER} definitions are applicable heavily depends on the tackled problem and applied network architecture; not all definitions are applicable in all cases.
We here propose a generalized model for computing a WER in MIMO situations that poses minimal requirements w.r.t. ancillary information: It does neither require speaker labels nor timing information.
It can consequently be used as a scoring tool for arbitrary ASR engines providing arbitrarily detailed information.
It treats an utterance as an atomic unit that always has to appear continuously in an \gls{ASR} system's output and ensures that the order of utterances uttered by each speaker is maintained.
We show that our MIMO WER definition is a generalization of many existing WER definitions.

Equally important to a sound definition of WER for MIMO is an efficient algorithm for computing it. 
Finding the matching between multiple input and output sequences is a combinatorial problem that can become intractable with a naive implementation.
In \cite{Sklyar2022_MultiTurnRNNTStreaming,Raj2022_ContinuousStreamingMultiTalker}, for example, it was stated that the \gls{ORC WER} could not be computed for a recording containing many utterances. 
An efficient algorithm for the evaluation of multi-output models was presented in \cite{Fiscus2006_MultipleDimensionLevenshtein}, and released under the name \verb!asclite!. 
It casts the alignment of a system output to multiple reference transcriptions as a multi-dimensional Levenshtein distance calculation that can be computed by dynamic programming. 
Their interpretation of the multi-dimensional Levenshtein distance, however, allows words to be transcribed on an arbitrary output channel, i.e. it is not counted as an error when an utterance is split over multiple channels (which the MIMO WER and many other WER definitions penalize).
This type of error, however, can render a transcription practically unusable.
Their tool additionally requires detailed temporal information about word begin and end times, which is not readily available for all of today's ASR approaches, e.g., End-to-End systems \cite{Kanda2022_StreamingMultiTalkerASR,Raj2022_ContinuousStreamingMultiTalker}.

We provide an efficient implementation of the MIMO WER that is also based on the multi-dimensional Levenshtein distance, but includes the constraint that an utterance must be matched consistently and uninterruptedly with one hypothesis channel.
It leads to an efficient implementation of the \gls{ORC WER} with a complexity that is polynomial in the number of utterances instead of exponential, as it was the case in \cite{Sklyar2022_MultiTurnRNNTStreaming}.

The contributions of this paper are as follows: (a) We propose a generalized WER for MIMO \gls{ASR} systems and present an efficient implementation that also leads to efficient implementations of other WER definitions, (b) we discuss its relationship to and pitfalls of existing WER definitions, and  (c) we release a software package for MIMO WER computations named \verb!MeetEval!\footnote{\url{https://github.com/fgnt/meeteval}}.

%
%

\pagebreak
\section{MIMO WER}
\glsreset{uWER}

Recently, transcription systems have emerged that handle complex recordings containing speech of multiple speakers or speech overlaps.
Such systems process multi-party speech and produce a hypothesis that can contain multiple channels, thus we here call them MIMO \gls{ASR}\footnote{MIMO here does not refer to multiple microphone inputs. This discussion holds for both single- and multi-microphone systems.}.
For such systems, an alignment between multiple references (e.g., one per speaker) and multiple hypothesis channels (not necessarily one per speaker) has to be found before a \gls{WER} can be computed.

A common way to compute a WER in such a case is utterance-wise evaluation, i.e., \gls{uWER} \cite{Yoshioka2018_RecognizingOverlappedSpeech,Chen2020_ContinuousSpeechSeparation,vonNeumann2021_GraphPITGeneralizedPermutation}.
A recording is here not evaluated in its entirety, but utterances are extracted from the input signal using ground truth utterance begin and end times before the \gls{ASR} system 
is applied.
For each utterance, the SISO WER is computed and the output channel is selected that minimizes the error.
%
Among others, this has the drawback that any text produced by the \gls{ASR} system outside the ground truth utterances is ignored, so the performance is constantly over-estimated.

For a more realistic assessment an evaluation is needed of the full hypothesis, also called continuous evaluation \cite{Chen2020_ContinuousSpeechSeparation}.
While an ideal MIMO \gls{ASR} system provides information about speaker identities and timing in addition to the transcription, i.e., answering the question \enquote{who said what and when?} \cite{Kanda2020_JointSpeakerCounting,Raj2021_IntegrationSpeechSeparation}, many systems focus only on sub-problems and neglect speaker identification \cite{Yoshioka2018_RecognizingOverlappedSpeech,Chen2020_ContinuousSpeechSeparation,Kanda2020_SerializedOutputTraining,Sklyar2022_MultiTurnRNNTStreaming} or temporal alignments \cite{Kanda2020_SerializedOutputTraining,Kanda2022_StreamingMultiTalkerASR}.
Many of the already proposed WER definitions for continuous evaluation (see \cref{sec:discussion} for discussion) are not applicable in all scenarios.


\subsection{A generalized WER definition for multi-speaker ASR}

We propose two natural assumptions for a WER model that satisfies all use cases stated before:
(i) The order of utterances produced by the same speaker cannot be changed by the \gls{ASR} system; and
(ii) a continuous portion of speech uttered by a single speaker without significant pauses should always appear continuously on one hypothesis channel. We call such a continuous portion of speech an utterance and a violation of assumption (ii) a channel switch.
Note that these two assumptions do not state any temporal ordering between utterances of different speakers.
We do not assume that a system estimates diarization or speaker information because we argue that the \gls{WER} should be applicable even if such information is not available.

Given $\nspk$ reference sequences of utterances $\tref_\ispk$ and $\nchn$ hypothesis channels $\thyp_\ichn$ as sequences of words, we find the assignment of reference utterances to hypothesis channels that minimizes the SISO \gls{WER} over all outputs, respecting the aforementioned assumptions (i) and (ii).
We call this error rate the MIMO WER.
It does only count transcription errors and no speaker attribution errors (in case speaker information is available).

Finding the MIMO assignment is computationally demanding as the number of valid assignments is exponential in $\nspk$, $\nchn$ and the number of utterances in the references.
The next section discusses how to find such an assignment for the MIMO WER by an extension of the Levenshtein distance.

\subsection{Efficient computation of the MIMO WER}
\label{sec:mimo-alg}

The Levenshtein distance \cite{Levenshtein1965_BinaryCodesCapable} between two word sequences $\tref$ and $\thyp$ can be efficiently computed with the Wagner-Fischer algorithm \cite{Wagner1974_StringtoStringCorrectionProblem}, which is a dynamic programming algorithm.  
With the indices $\iref$ and $\ihyp$, representing indices into the word sequences $\tref$ and $\thyp$, respectively,
a two-dimensional distance matrix $\lm$ is filled as follows: starting the recursion with 
$\lm(0, h)=h$ and $\lm(r, 0)=r$ the entries of the matrix are recursively computed as
\begin{align}
    \lm(\iref, \ihyp) &= \min
    \begin{cases}
    \lm(\iref - 1, \ihyp - 1) + \costcorrsub \\
    \lm(\iref, \ihyp - 1) + \costins\\
    \lm(\iref-1, \ihyp) + \costdel,
    \end{cases} \label{eq:lev}
\end{align} 
where $\costcorrsub,\costins$ and $\costdel$ are the costs of a correct match or substitution, an insertion, and a deletion operation, respectively.
After $\lm$ has been filled, the Levenshtein distance between the two word sequences is given by $\lm(|\tref|,|\thyp|)$, where $|\cdot|$ denotes the length of the sequence. 
The costs for substitution or deletion are $\costcorrsub=\costcorr$ if $\tref(\iref)=\thyp(\ihyp)$ and $\costcorrsub=\costsub$ otherwise.
%
While for a correct recognition $\tref(\iref)=\thyp(\ihyp)$ the transition cost is zero ($\costcorr=0$), substitutions, insertions and deletions incur costs of typically $(\costsub, \costins, \costdel) = (1,1,1)$, while other values have also been used \cite{Povey2011_KaldiSpeechRecognition,Fiscus2006_MultipleDimensionLevenshtein}.
The optimal alignment between reference and hypothesis can be found by a backtracking pass through $\lm$, starting from $(|\tref|,|\thyp|)$ and moving along the path given by the transitions that achieved the minimum in \cref{eq:lev},  until ending in $\lm(0,0)$.

While a brute force computation of all possible alignments between $\tref$ and $\thyp$ would have a complexity that is exponential in the number of words, this dynamic programming algorithm has a complexity that grows only linearly with the length of the word sequences. 
Its complexity is given by $\bigo(3|\tref||\thyp|)$ where the factor of 3 is the number of computations required to obtain one matrix element, see \cref{eq:lev}.

\subsubsection{Multi-dimensional Levenshtein distance}
Moving from the above single reference and single hypothesis case to $\nspk$ references and $\nchn$ hypothesis channels $\tref_1,...,\tref_\nspk$ and $\thyp_1,...,\thyp_\nchn$, the computational complexity can again be dramatically reduced from a brute-force search by extending the two-dimensional  Levenshtein distance matrix to a multi-dimensional tensor. 
The indices into $\lm$ are extended to multi-indices $\mref =(\iref_1,\ldots , \iref_{\nspk})$ and $\mhyp = (\ihyp_1, \ldots , \ihyp_{\nchn})$, where $\iref_\ispk$ indexes $\tref_\ispk$ and $\ihyp_\ichn$ indexes $\thyp_\ichn$.
If we first neglect assumption (ii) of the MIMO WER, the distance tensor is recursively filled by:
\begin{align}
    \lm(\mref,\mhyp) =
    \min_{\substack{\aref\in\{1,\ldots, \nspk\} \\ \ahyp\in\{1, \ldots, \nchn\}}}
        \begin{cases}
        \lm(\mref-\uv_\aref^{(\nspk)},\mhyp-\uv_\ahyp^{(\nchn)}) + \costcorrsub \\
        \lm(\mref-\uv_\aref^{(\nspk)},\mhyp) + \costdel \\
        \lm(\mref,\mhyp-\uv_\ahyp^{(\nchn)}) + \costins, \\
        \end{cases}
    \label{eq:lev-multi-dimensional}
\end{align}
where $\munit_\aref^{(\nspk)}$ is an $\nspk$-dimensional multi-index containing all zeros, except for a one in the $\aref$-th position. Similarly, $\munit_\ahyp^{(\nchn)}$ is $\nchn$-dimensional with a one on the $\ahyp$-th component.

\cref{eq:lev-multi-dimensional} is equivalent to the derivations in \cite{Fiscus2006_MultipleDimensionLevenshtein} for simple word strings and simplifies to \cref{eq:lev} for $\nspk=\nchn=1$.
Filling $\lm$ with \cref{eq:lev-multi-dimensional} requires  $\bigo((\nspk\nchn + \nspk + \nchn)(\prod_\aref^\nspk |\tref_\aref|) (\prod_\ihyp^\nchn |\thyp_\ihyp|))$ steps, which is exponential in the number of references $\nspk$ and hypothesis channels $\nchn$. 
Here, the number of computations to obtain one element of $\lm$ is given by $\nspk\nchn$, the number of substitutions, plus $ \nspk$ deletions plus $\nchn$ insertions.
This formulation implies the first MIMO assumption that the order of utterances produced by the same speaker must not change, but it does not include the second assumption that an utterance should be consistent on an output.

\begin{figure*}[t]
    \centering
    \begin{subfigure}{17em}
        \centering
        \input{tikz/example_wrong_diarization}%
        \caption{cpWER counts diarization errors while speaker-agnostic WERs do not.}
        \label{fig:ex-incorrect-diarization}
    \end{subfigure}
    \hspace{1em}
    \begin{subfigure}{17em}
        \centering
        \input{tikz/example_asclite_alternating_words}%
        \cprotect\caption{\verb!asclite! does not penalize channel switches within an utterance.}
        \label{fig:ex-asclite}
    \end{subfigure}
    \hspace{1em}
    \begin{subfigure}{17em}
        \centering
        \input{tikz/example_orc_wrong_annotation}%
        \caption{ORC WER can over-estimate the WER when annotations are faulty.}
        \label{fig:ex-orc-wer}
    \end{subfigure}
    \cprotect\caption{Toy examples to demonstrate differences of WER definitions. 
    Each box 
    \tikz[baseline=-0.75ex]{\node[a]{a};}
    represents a word and a light gray box
    \tikz[baseline=-0.75ex]{\node[utterance]{};} represents an utterance.
    Errors counted by the different WER definitions are given in the tables.
}\label{fig:examples}%
    \vspace{-2em}
\end{figure*}
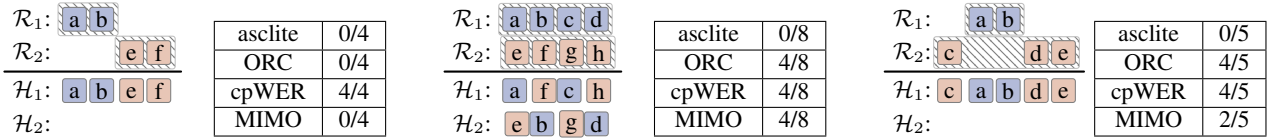%

\subsubsection{Proposed: MIMO Levenshtein distance}
In order to include utterance splits in the error count, i.e., assumption (ii), we introduce a channel change token \changetoken{} that is inserted between utterances in all reference word sequences $\tref$.
Updates across references and hypotheses are only allowed at these change tokens.
We represent the currently active mapping of reference to hypothesis channel by a pair $(\aref,\ahyp)$ where reference $\tref_\aref$ is matched to hypothesis channel $\thyp_\ahyp$. 
The constrained update equation to compute the MIMO WER is given by (neglecting the superscript of the multi-index):
\begin{align}
    \lm_{\aref,\ahyp}(\mref,\mhyp) = \begin{cases}
        \displaystyle\min_{\substack{\tilde{\aref}\in\{1,\ldots, \nspk\} \\ \tilde{\ahyp}\in\{1, \ldots, \nchn\}}}\lm_{\tilde{\aref},\tilde{\ahyp}}(\mref-\munit_{\tilde{\aref}},\mhyp); \quad \text{if } \tref_\aref(\iref_\aref) = \rlap{\text{\changetoken}} \\
        \min\begin{cases}
            \lm_{\aref,\ahyp}(\mref-\munit_\aref,\mhyp-\munit_\ahyp) + \costcorrsub\\
            \lm_{\aref,\ahyp}(\mref-\munit_\aref,\mhyp) + \costdel \\ 
            \lm_{\aref,\ahyp}(\mref,\mhyp-\munit_\ahyp) + \costins .
        \end{cases}
    \end{cases}
    \label{eq:lev-mimo}
\end{align}
The lower $\min$ operation is equal to the two-dimensional string matching in \cref{eq:lev} along the slice of $\lm$ determined by $\aref$ and $\ahyp$ which allows leveraging efficient algorithms for Levenshtein distance computation, such as \cite{Myers1999_FastBitvectorAlgorithm,Masek1980_FasterAlgorithmComputing}, also in the multi-dimensional case.
\cref{eq:lev-mimo} collapses to \cref{eq:lev-multi-dimensional} if every second word in every reference $\tref_\aref$ is \changetoken.
Using \cref{eq:lev-mimo} causes $\lm$ to be relatively sparse in $\mref$ which slightly reduces the complexity compared to \cref{eq:lev-multi-dimensional}.
When choosing $\mref$ as a scalar ($\nspk=1$), this formulation allows for an efficient computation of the ORC WER (see \cref{sec:orcwer}).

Computing the MIMO or ORC WER is NP-hard for arbitrary $\nspk$ and $\nchn$.
This can be proven by showing that the string MERGE problem, which is NP-complete \cite{Mansfield1983_ComputationalComplexityMerge}, can be solved by checking if the MIMO WER for specifically arranged inputs is zero.
The MIMO assignment must be at least as hard as MERGE, thus, NP-hard.

    

\begin{table}[t]
    \centering
    \caption{
        Comparison of continuous evaluation WERs.
    }\label{tab:wer-comparison}
    \setlength{\tabcolsep}{1.5pt}
    \begin{tabular}{lcccc}
        \toprule
        & \multicolumn{1}{c}{speaker-attributed} & \multicolumn{3}{c}{speaker-agnostic}\\
        \cmidrule(r{.2em}){2-2}\cmidrule(l{.2em}){3-5}
         & cpWER & asclite & ORC & MIMO \\
         \midrule
         Penalizes \\
         \tbli Speaker Confusion & \cmark & \xmark & \xmark & \xmark \\
         \tbli Channel Switches & \cmark & \xmark & \cmark & \cmark \\
         \tbli Temporal Errors & \xmark & \cmark & (\cmark)\footnotemark[3] & \xmark \\
         \midrule
         Examples (applicable to) \\
         \tbli CSS + Diar. \cite{Raj2021_IntegrationSpeechSeparation} & \cmark & (\cmark)\footnotemark[4] & (\cmark)\footnotemark[4] & (\cmark)\footnotemark[4] \\
         \tbli Diar. + Sep. + ASR \cite{Watanabe2020_CHiME6ChallengeTackling} & \cmark & (\cmark)\footnotemark[4] & (\cmark)\footnotemark[4] & (\cmark)\footnotemark[4] \\
         \tbli CSS + ASR ($\nchn<\nspk$) \cite{Chen2020_ContinuousSpeechSeparation} & \xmark & \cmark & \cmark & \cmark \\
         \tbli SOT ($\nchn=1$) \\
         \tblii FIFO \cite{Kanda2020_SerializedOutputTraining,Kanda2022_StreamingMultiTalkerASR} & \xmark & \xmark & \cmark & \cmark \\
         \tbliii + speaker ID \cite{Kanda2020_JointSpeakerCounting} & \cmark & \xmark & \cmark & \cmark \\
         \tblii non-FIFO \cite{Kanda2020_SerializedOutputTraining} & \xmark & \xmark & \xmark & \cmark \\ 
         \tbli {\footnotesize\tab{ MT-RNN-T\cite{Sklyar2021_StreamingMultiSpeakerASR}\\ SURT\cite{Raj2022_ContinuousStreamingMultiTalker}}} ($\nchn<\nspk$) & \xmark & (\cmark)\footnotemark[5] & \cmark & \cmark \\ 
         \bottomrule
    \end{tabular}
    \label{tab:my_label}
\end{table}
\footnotetext[3]{ORC WER only penalizes temporal errors when the order of utterances is changed by them.}
\footnotetext[4]{Applicable in theory, but temporal or memory complexity may explode for large numbers of output channels.}
\footnotetext[5]{This architecture may or may not provide temporal information.}

\section{Discussion}
\label{sec:discussion}

\cref{tab:wer-comparison} gives a short overview of the presented WER definition and shows example use-cases where each one is applicable. \cref{fig:examples} displays small scoring examples to highlight their differences.

\subsection{Speaker attributed WERs}
\glsreset{SA-WER}


It is common to compute a \gls{SA-WER} \cite{nist_2009_2009,Watanabe2020_CHiME6ChallengeTackling} when an \gls{ASR} system provides speaker labels, e.g., for \gls{SA-ASR} \cite{Kanda2020_JointSpeakerCounting}.
\gls{SA-WER} judges both, transcription and speaker attribution errors.
The \gls{SA-WER} definition from \cite{nist_2009_2009} can be computed when the estimated speaker labels represent the true speaker identity, i.e., the mapping between estimated speaker labels and reference labels is known.
%
In recent publications, the \gls{cpWER} \cite{Watanabe2020_CHiME6ChallengeTackling} is preferred over the \gls{SA-WER}.
It is the overall SISO \gls{WER} across all channels for the permutation of reference speakers $|\tref_\ispk|$ to hypothesis speakers $|\thyp_\ichn|$ with the minimal error.

The \gls{cpWER} is a special case of the MIMO model with the constraint that the mapping between reference speakers $|\tref_\ispk|$ and hypotheses $|\thyp_\ichn|$ is bijective, i.e., a permutation in $\tref_\ispk$ and $\nchn=\nspk$.
Under-estimation (i.e., when $\nchn<\nspk$) is handled by adding empty dummy channels until $\nchn=\nspk$, and over-estimation is handled implicitly by having hypothesis channels that are not matched with any reference.
The \gls{cpWER} is an upper bound on the MIMO WER since it reduces the number of assignments.
It can be computed in polynomial time with the Hungarian algorithm \cite{Kuhn1955_HungarianMethodAssignment,Munkres1957_AlgorithmsAssignmentTransportation,vonNeumann2020_SpeedingPermutationInvariant}.
An example for the \gls{cpWER} judging speaker attribution errors is shown in \cref{fig:ex-incorrect-diarization}.

\subsection{Speaker agnostic WERs}

If speaker information is not available, e.g., for \gls{CSS} systems \cite{Chen2020_ContinuousSpeechSeparation}, a speaker-agnostic \gls{WER} has to be computed.
A speaker-agnostic WER can also be useful when speaker information is available as the difference between a speaker-attributed and a speaker-agnostic \gls{WER} indicates how many errors can be attributed to mistakes regarding diarization.

\subsubsection{asclite}
A widely used tool for computing a speaker-agnostic WER is the \verb!asclite! tool \cite{Fiscus2006_MultipleDimensionLevenshtein}, as used, e.g., in \cite{Chen2020_ContinuousSpeechSeparation,Kanda2020_SerializedOutputTraining}.
It uses a multi-dimensional Levenshtein distance algorithm (see \cref{eq:lev-multi-dimensional}) to match multiple references to multiple hypothesis channels.
It reduces the search space significantly by use of timestamps for words in the references and hypothesis
.
It has, as such, the tightest bounds on estimated timestamps among the discussed WER definitions and is the only WER that penalizes time annotation errors.
This adds the constraint that \verb!asclite! can only be used when the \gls{ASR} system produces such timing information, which is not available for all methods \cite{Kanda2022_StreamingMultiTalkerASR}.
It can, nevertheless, have an exploding complexity in certain cases even when temporal alignment information is available  \cite{Kanda2022_VarArrayMeetsTSOT}.
The idea of \verb!asclite! is similar to our MIMO WER when assumption (ii) is dropped, i.e., utterance splits are not counted as an error.
It thus produces overoptimistic error rates. 
An extreme example of this is visualized in \cref{fig:ex-asclite}, where words of an utterance are spread over multiple channels which renders the hypothesis practically unusable, while \verb!asclite! computes a WER of zero.
All other presented WER definitions respect MIMO assumption (ii).

\subsubsection{ORC WER}
\label{sec:orcwer}

The \gls{ORC WER} has been proposed \cite{Sklyar2022_MultiTurnRNNTStreaming} as a speaker-agnostic WER that does not judge temporal alignment but respects MIMO assumption (ii).
It is a special case of our MIMO model that merges all $\nspk$ references into one reference $\treforc$, sorted by utterance begin times, and then computes the MIMO WER using only $\treforc$. 
This is equal the constraint that the global order of utterances must not change.
It can thus not be used when the system is allowed to change the ordering of the utterances, as, e.g., a non-FIFO SOT \cite{Kanda2020_SerializedOutputTraining} does.
Such a re-ordering can also happen implicitly, e.g., when the given timestamps are faulty or ambiguous.
\cref{fig:ex-orc-wer} shows an example of such a corner case where two utterances are wrongly labeled as a single utterance and the \gls{ORC WER} counts all words as substitutions while the \gls{ASR} system transcribed everything correctly.
Such a constellation is, however, often only a corner-case that does not appear frequently in common evaluation scenarios.

The ORC WER is a lower bound on the \gls{cpWER} since it does not consider speaker errors, and becomes equal to \gls{cpWER} if no speaker errors are present.
It is at the same time an upper bound on the MIMO WER and becomes equal to the MIMO WER when the utterance ordering is unambiguous and the number of errors is small.

The naive implementation in \cite{Sklyar2022_MultiTurnRNNTStreaming} is computationally expensive with a complexity that is exponential in the number of utterances, which caused the authors of \cite{Sklyar2022_MultiTurnRNNTStreaming} to drop examples exceeding 23 utterances for the WER evaluation.
The complexity can be reduced to being polynomial in the number of utterances by using \cref{eq:lev-mimo} with $\nspk=1$.
It thus allows computing the ORC WER for longer recordings (see \cref{sec:benchmark}).

\section{Benchmark}
\label{sec:benchmark}

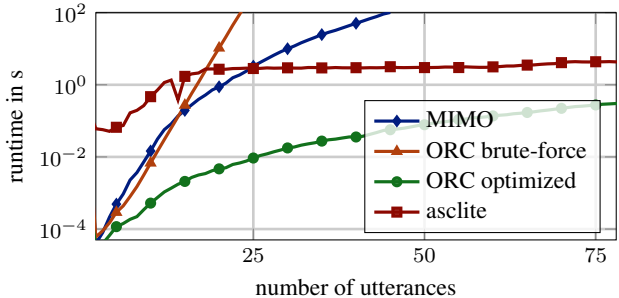
\begin{figure}[t]
    \centering
    \input{tikz/runtime_plot_num_utterances}
    \caption{
    Runtime comparison of different WERs for a CSS scenario with $\nspk = 4$ speakers and $\nchn = 2$ output channels. 
    }
    \label{fig:runtime-num-utterances}
\end{figure}
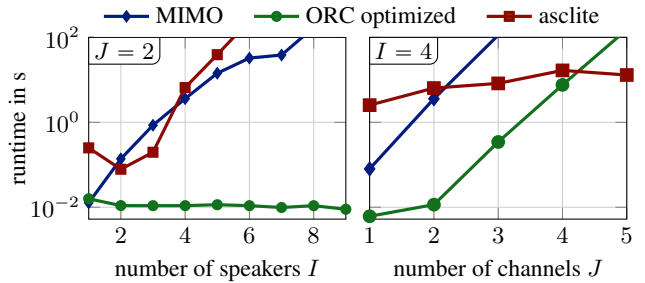
\begin{figure}[t]
    \centering
    \input{tikz/runtime_plot_num_speakers_channels}
    \caption{
    Runtime comparison for different numbers of speakers $\nspk$ and output channels $\nchn$.
    The number of utterances is 25 in both plots.
    }
    \label{fig:runtime-num-speakers}
\end{figure}


\cref{fig:runtime-num-utterances} show the runtime of the different \glspl{WER} algorithms over the number of utterances to be scored. 
The benchmark was run on a single core of an AMD Milan 7763 CPU with 2.45GHz in the Noctua2 compute cluster of the Paderborn Center for Parallel Computing (PC$^2$).
We simulated a common \gls{CSS} scenario with four reference speakers and two system output streams \cite{Chen2020_ContinuousSpeechSeparation}.
The \gls{cpWER} is excluded from the plot because it perform no complex alignment and thus has a much lower complexity compared to the other WERs.
The measurements for \verb!asclite! contain a slight offset due to file parsing overhead since no Python interface is available.
Note that \verb!asclite! appears very efficient in this benchmark because it uses temporal annotations to narrow down the search space.
We here evaluate the MIMO WER in its most general case, i.e., without using temporal information.
The MIMO WER can be tuned to use the same information as asclite with a similar complexity and thus runtime.

From \cref{fig:runtime-num-utterances} we can see that the MIMO and ORC WERs indeed have a complexity that is polynomial in the number of utterances (concavity of the curve in log scale).
The complexity of MIMO WER is greater than that of ORC WER because ORC WER removes the depenency on the number of speakers $\nspk$.
The brute-force implementation of the \gls{ORC WER} explodes for about 20 utterances, as already reported by \cite{Sklyar2022_MultiTurnRNNTStreaming}.
\verb!asclite! shows a behavior that is close to linear in this scenario.

\cref{fig:runtime-num-speakers} shows the dependency on the number of speakers and the number of channels, respectively.
The complexity of the MIMO WER and \verb!asclite! is exponential in the number of speakers $\nspk$ while the \gls{ORC WER} has a constant complexity w.r.t. the number of speakers since the references are merged.
Being independent of the number of speakers $\nspk$ is important for evaluating \gls{CSS} systems where the number of speakers $\nspk$ can be arbitrarily large but the number of output channels $\nchn$ is small.
The complexity of \verb!asclite! explodes in more complex scenarios, such as generated by the \verb!mms_msg! tool \cite{Cord-Landwehr2022_MMSMSGMultipurposeMultispeaker}.
Here, the only known applicable tool remains our proposed efficient \gls{ORC WER} implementation.

We can conclude that, while the \gls{ORC WER} has a few corner cases, it is usually well suited for evaluation with a reasonable execution time.
Systems that are incompatible to \gls{ORC WER} (e.g., non-FIFO SOT) can utilize the MIMO framework and open-source implementation for an efficient WER computation.



\section{Conclusion}
\addtocounter{footnote}{-2}%
\addtocounter{Hfootnote}{-2}%
We proposed MIMO WER, a generalized WER definition to assess modern ASR systems that transcribe multiple speakers' utterances to multiple output channels.
We embedded it in a comprehensive discussion of existing \glspl{WER} showing that the MIMO WER can cater a wide range of applications, and that other WER definitions are specialization for particular use cases. 
An efficient implementation based on a multi-dimensional Levenshtein distance definition is derived for MIMO and ORC, and a channel change token is introduced to penalize the split of utterances over different ASR output channels.
The software is provided as an open source tool \verb!MeetEval!\footnotemark.


\section{Acknowledgement}

Christoph Boeddeker was funded by Deutsche Forschungsgemeinschaft (DFG), project no.\ 448568305. 
Computational resources were provided by the Paderborn Center for Parallel Computing.

\vfill\pagebreak

\bibliographystyle{IEEEbib}
\balance
\bibliography{references.bib}

\end{document}

%% file: macros.tex
\usepackage{amsmath}
\usepackage{amssymb}
\usepackage{mathabx}
\usepackage{bbm}
\usepackage{physics}
\usepackage{siunitx}
\usepackage{trfsigns}
\usepackage{pifont}
\usepackage{etoolbox}
\usepackage{multirow}
\usepackage{csquotes}
\usepackage[shortlabels]{enumitem}  
\usepackage{cite}   
\usepackage{hyperref}
\usepackage[capitalize]{cleveref}
\usepackage{xifthen}
\usepackage{balance} 
\usepackage{booktabs}
\usepackage{listings}
\usepackage{listofitems}
\usepackage{subcaption}
\usepackage{cprotect}  
\usepackage{sepfootnotes} 

\usepackage{url}

\newcommand{\cmark}{\ding{51}}
\newcommand{\xmark}{{--}}

\let\oldcite\cite
\renewcommand{\cite}[1]{%
\ifthenelse{\isempty{#1}}%
{\inred{[cite!]}}%
{\oldcite{#1}}%
}

\newcolumntype{H}{>{\setbox0=\hbox\bgroup}c<{\egroup}@{}}   

\newcommand{\tab}[2][c]{\begin{tabular}{@{}#1@{}}#2\end{tabular}}   

\newcommand{\tbli}{\phantom{+}}
\newcommand{\tblii}{\phantom{++}}
\newcommand{\tbliii}{\phantom{+++}}

\newcommand*{\thl}{\fontseries{b}\selectfont}
\robustify\thl
\robustify\fontseries

\newcommand{\inred}[1]{\textcolor{red}{#1}}

\makeatletter
\renewcommand\subsubsection{\@startsection{subsubsection}{3}{\z@}%
                                     {-1.25ex\@plus -.2ex \@minus -.2ex}%
                                     {0.5ex \@plus .2ex}%
                                     {\normalfont\normalsize\bfseries}}
\makeatother

%% file: tikz_style.tex
\usepackage{tikz}
\usepackage{pgfplots}
\pgfplotsset{compat=1.16}
\usepgfplotslibrary{groupplots}
\usetikzlibrary{positioning,arrows,matrix,fit,calc,patterns,chains,scopes,shapes.multipart,decorations,decorations.markings,backgrounds}

\definecolor{palette-1}{HTML}{001C7F}    
\definecolor{palette-2}{HTML}{B1400D}    
\definecolor{palette-3}{HTML}{12711C}    
\definecolor{palette-4}{HTML}{8C0800}    
\definecolor{palette-5}{HTML}{591E71}    
\definecolor{palette-6}{HTML}{592F0D}    
\definecolor{palette-7}{HTML}{A23582}    
\definecolor{palette-8}{HTML}{3C3C3C}    
\definecolor{palette-9}{HTML}{B8850A}    
\definecolor{palette-10}{HTML}{006374}      

\tikzset{
    line/.style={draw,black,thick,rounded corners=1mm,line cap=round},
    noshortarrow/.style={line,->},
    arrow/.style={noshortarrow,shorten >=.3mm},
    doublearrow/.style={arrow,<->, shorten <=.3mm},
    box/.style={draw,black,thick,minimum height=3em,text height=1.5ex,text depth=0.25ex,rounded corners=3,fill=white},
    nopadding/.style={minimum height=0,inner sep=1mm},
    signalbox/.style={draw,black,thin,rounded corners=1mm,minimum width=7mm, minimum height=4mm,inner sep=0},
    pbox/.style={box,fill=black!10},
    backgroundbox/.style={inner xsep=3mm, inner ysep=1mm, draw, dashed, rounded corners,fill=orange!10},
    branch/.style={inner sep=0.3mm,circle,fill=black},
    operator/.style={draw,circle,black,rounded corners,inner sep=0,fill=white},
    vertex/.style={draw,ultra thin,circle,black,rounded corners,inner sep=0.6mm,fill=gray,fill opacity=0.5},
    edge/.style={line,very thick,line cap=butt},
    pattern1/.style={pattern=north west lines,pattern color=palette-1},
    pattern2/.style={pattern=north east lines,pattern color=palette-2},
    pattern3/.style={pattern=crosshatch,pattern color=palette-3},
    buswidth/.style={path picture={\draw[black,-] (path picture bounding box.south west) -- (path picture bounding box.north east);}},
    plotlabel/.style={inner sep=0.2em, append after command={
   \pgfextra
        \path[sharp corners, fill=white]%
    (\tikzlastnode.north east)[rounded corners=3pt,draw]--(\tikzlastnode.south east) [sharp corners,draw]-- (\tikzlastnode.south west)[sharp corners] |- cycle;
   \endpgfextra}}
}

\tikzset{
    word/.style={draw=gray,inner sep=1,rounded corners=1,minimum height=1em,minimum width=1em,font=\vphantom{A}},
    a/.style={word,fill=palette-1!30},
    b/.style={word,fill=palette-2!30},
    c/.style={word,fill=palette-3!30},
    d/.style={word,fill=palette-4!30},
    e/.style={word,fill=palette-5!30},
    f/.style={word,fill=palette-6!30},
    g/.style={word,fill=palette-7!30},
    h/.style={word,fill=palette-8!30},
    i/.style={word,fill=palette-9!30},
    k/.style={word,fill=palette-10!30},
    utterance/.style={word,draw=lightgray,fill=lightgray!10,pattern=north west lines,pattern color=gray,inner sep=1.5},
    channel/.style={minimum width=1.6cm, minimum height=1.2em},
    matchline/.style={line,thin},
    label/.style={inner sep=0,font=\scriptsize\sffamily, anchor=south west,gray},
    deletion/.style={word,color=lightgray,dotted,fill=lightgray!10},
    insertion/.style={very thick, dashed, draw=red}
}

%% file: symbols.tex
\newcommand\defm[2]{\expandafter\newcommand{#1}{\ensuremath{#2}}} 

\defm\Loss{\mathcal{L}}    

\newcommand{\vect}[1]{\ensuremath{\boldsymbol{\mathbf{#1}}}}

\defm\bigo{\mathcal{O}}
\defm\naturals{\mathbb{N}}

\defm\tref{\mathcal{R}}
\defm\thyp{\mathcal{H}}
\defm\treforc{\tref^\text{(ORC)}}

\defm\iref{r}
\defm\ihyp{h}
\defm\ispk{i}
\defm\ichn{j}

\defm\mref{\vect{r}}
\defm\mhyp{\vect{h}}
\defm\munit{\vect{e}}

\defm\nspk{I}
\defm\nchn{J}
\defm\nref{R} 

\defm\aref{i}
\defm\ahyp{j}

\defm\costcorr{C_\text{corr}}
\defm\costsub{C_\text{sub}}
\defm\costdel{C_\text{del}}
\defm\costins{C_\text{ins}}
\defm\costcorrsub{C_\text{corr/sub}}

\defm\changetoken{\text{\textless ct\textgreater}}

\defm\lm{L}
\defm\li{r}
\defm\lj{h}

\defm\mi{\vect{i}}
\defm\mj{\vect{j}}
\defm\lmd{\lm}
\defm\ri{r}     
\defm\riv{\vect{r}} 
\defm\hi{h}     
\defm\hiv{\vect{h}} 
\defm\uv{\vect{e}}  
\defm\ai{k}   
\defm\bi{c}   


%% file: glossaries.tex
\usepackage[acronym,shortcuts]{glossaries}
\glsdisablehyper    

\newacronym{ASR}{ASR}{Automatic Speech Recognition}
\newacronym{CSS}{CSS}{Continuous Speech Separation}
\newacronym{SOT}{SOT}{Serialized Output Training}
\newacronym{WER}{WER}{Word Error Rate}
\newacronym{ORC WER}{ORC WER}{Optimal Reference Combination WER}
\newacronym{cpWER}{cpWER}{Concatenated Minimum Permutation WER}
\newacronym{uWER}{uWER}{utterance-wise WER}
\newacronym{SA-WER}{SA-WER}{Speaker Attributed WER}
\newacronym{SA-ASR}{SA-ASR}{Speaker Attributed ASR}

%% file: tikz/example_wrong_diarization.tex
\begin{tikzpicture}[node distance=0.1em]

    \coordinate (origin) at (0, 0);
    \coordinate (ref1) at ($(origin) + (0,-1em)$);
    \coordinate (ref2) at ($(ref1) + (0,-1.4em)$);
    \coordinate (hyp1) at ($(ref2) + (0,-1.6em)$);
    \coordinate (hyp2) at ($(hyp1) + (0,-1.4em)$);
    
    {[start chain=going right]
        \node[a,on chain,anchor=west] (a) at (ref1) {a};
        \node[a,on chain] (b) {b};
    }
    \begin{pgfonlayer}{background}
        \node[utterance,fit={(a)(b)}]{};
    \end{pgfonlayer}
    
    {[start chain=going right]
        \node[b,on chain,xshift=2.3em] (e) at (ref2) {e};
        \node[b,on chain] (f) {f};
    }
    \begin{pgfonlayer}{background}
        \node[utterance,fit={(e)(f)}] (utterance2){};
    \end{pgfonlayer}

    \node[channel, anchor=west,xshift=-0.1em] at (hyp1) (channel1) {};
    \node[channel, anchor=west,xshift=-0.1em] at (hyp2) (channel2) {};
        \node[a] at (a|-channel1) (ha) {a};
        \node[a,] at (b|-channel1) (hb) {b};
        \node[b] at (e|-channel1) (he) {e};
        \node[b] at (f|-channel1) (hf) {f};

    \node[left=0.1em of ref1] {$\mathcal{R}_1$:};
    \node[left=0.1em of ref2] {$\mathcal{R}_2$:};
    \node[left=0.1em of hyp1] {$\mathcal{H}_1$:};
    \node[left=0.1em of hyp2] (hyp2) {$\mathcal{H}_2$:};
    
    \coordinate (divider) at ($(ref2)!0.5!(hyp1)$);
    \draw[line] (hyp2.west|-divider) -- (channel1.east|-divider);
    
    \node[xshift=10em,yshift=-3.5em] {
        \begin{tabular}{|c|c|}
            \hline
            asclite & 0/4 \\
            \hline
            ORC & 0/4 \\
            \hline
            cpWER & 4/4 \\
            \hline
            MIMO & 0/4 \\
            \hline
        \end{tabular}
    };    
\end{tikzpicture}

%% file: tikz/example_asclite_alternating_words.tex
\begin{tikzpicture}[node distance=0.1em]

    \coordinate (origin) at (0, 0);
    \coordinate (ref1) at ($(origin) + (0,0em)$);
    \coordinate (ref2) at ($(ref1) + (0,-1.4em)$);
    \coordinate (hyp1) at ($(ref2) + (0,-1.6em)$);
    \coordinate (hyp2) at ($(hyp1) + (0,-1.4em)$);
    
    {[start chain=going right,local bounding box=refutt1]
        \node[a,on chain,anchor=west] (a) at (ref1) {a};
        \node[a,on chain] (b) {b};
        \node[a,on chain] (c) {c};
        \node[a,on chain] (d) {d};
    }
    \begin{pgfonlayer}{background}
        \node[utterance,fit={(refutt1)}]{};
    \end{pgfonlayer}
    
    {[start chain=going right,local bounding box=refutt2]
        \node[b,on chain] (e) at (ref2) {e};
        \node[b,on chain] (f) {f};
        \node[b,on chain] (g) {g};
        \node[b,on chain] (h) {h};
    }
    \begin{pgfonlayer}{background}
        \node[utterance,fit={(e)(h)}]{};
    \end{pgfonlayer}

    \node[channel, anchor=west,xshift=-0.1em] at (hyp1) (channel1) {};
    \node[channel, anchor=west,xshift=-0.1em] at (hyp2) (channel2) {};
        \node[a] at (a|-channel1) (ha) {a};
        \node[a,] at (b|-channel2) (hb) {b};
        \node[a,] at (c|-channel1) (hc) {c};
        \node[a,] at (d|-channel2) (hd) {d};
        \node[b,] at (e|-channel2) (he) {e};
        \node[b,] at (f|-channel1) (hf) {f};
        \node[b,] at (g|-channel2) (hg) {g};
        \node[b,] at (h|-channel1) (hh) {h};

    \node[left=0.1em of ref1] {$\mathcal{R}_1$:};
    \node[left=0.1em of ref2] {$\mathcal{R}_2$:};
    \node[left=0.1em of hyp1] {$\mathcal{H}_1$:};
    \node[left=0.1em of hyp2] (hyp2) {$\mathcal{H}_2$:};
    
    \coordinate (divider) at ($(ref2)!0.5!(hyp1)$);
    \draw[line] (hyp2.west|-divider) -- (channel1.east|-divider);
    
    \node[xshift=10em,yshift=-2.5em] {
        \begin{tabular}{|c|c|}
        \hline
        asclite & 0/8 \\
        \hline
        ORC & 4/8 \\
        \hline
        cpWER & 4/8 \\
        \hline
        MIMO & 4/8 \\
        \hline
    \end{tabular}
    };    
\end{tikzpicture}

%% file: tikz/example_orc_wrong_annotation.tex
\begin{tikzpicture}[node distance=0.1em,]
    {[local bounding box=out]
    \coordinate (origin) at (0, 0);
    \coordinate (ref1) at ($(origin) + (0,0)$);
    \coordinate (ref2) at ($(ref1) + (0,-1.4em)$);
    \coordinate (hyp1) at ($(ref2) + (0,-1.6em)$);
    \coordinate (hyp2) at ($(hyp1) + (0,-1.4em)$);
    
    {[local bounding box=ref]
        {[local bounding box=spk1]
            {[start chain=going right, local bounding box=utt1]
                \node[a,on chain,anchor=west,right=1.1em of ref1] (a) {a};
                \node[a,on chain] (b) {b};
            }
            \begin{pgfonlayer}{background}
                \node[utterance,fit={(utt1)}]{};
            \end{pgfonlayer}
        }
        
        {[local bounding box=spk2]
            {[start chain=going right,local bounding box=utt2]
                \node[b,on chain,xshift=0.3em] (c) at (ref2) {c};
                \node[b,on chain, xshift=2.5em] (d) {d};
                \node[b,on chain] (e) {e};
            }
            \begin{pgfonlayer}{background}
                \node[utterance,fit={(utt2)}] (utterance2){};
            \end{pgfonlayer}
        }
    }

    \node[channel, anchor=west,xshift=-0.1em,minimum width=1.9cm] at (hyp1) (channel1) {};
    \node[channel, anchor=west,xshift=-0.1em] at (hyp2) (channel2) {};
        \node[a] at (a|-channel1) (ha) {a};
        \node[a,] at (b|-channel1) (hb) {b};
        \node[b,] at (c|-channel1) (hc) {c};
        \node[b,] at (d|-channel1) (hd) {d};
        \node[b,] at (e|-channel1) (he) {e};

    \node[left=0.1em of ref1] {$\tref_1$:};
    \node[left=0.1em of ref2] {$\tref_2$:};
    \node[left=0.1em of hyp1] {$\thyp_1$:};
    \node[left=0.1em of hyp2] (hyp2) {$\thyp_2$:};
    
    \coordinate (divider) at ($(ref2)!0.5!(hyp1)$);
    \draw[line] (hyp2.west|-divider) -- (channel1.east|-divider);
    \newcommand{\zero}{\color{gray!50}0}
    \node[xshift=10em,yshift=-2.5em] {
        \begin{tabular}{|c|c|}
        \hline
        asclite & 0/5 \\
        \hline
        ORC & 4/5\\
        \hline
        cpWER & 4/5 \\
        \hline
        MIMO & 2/5 \\
        \hline
    \end{tabular}
    };%
    }%
\end{tikzpicture}%

%% file: tikz/runtime_plot_num_utterances.tex
\begin{tikzpicture}
\begin{axis}[
    xlabel=number of utterances,
    ylabel=runtime in s,
    xmin=2, xmax=78,
    ymin=0.00005, ymax=100,
    ymode=log,
    log basis y={10},
    grid=both,
    ytick={1e-4,1e-2,1,1e2},
    grid style={line width=1pt,draw=black!20},
    legend cell align={left},
    legend style={
      fill opacity=0.8,
      draw opacity=1,
      text opacity=1,
      at={(0.97,0.03)},
      anchor=south east,
    },
    xtick={25,50,75},
    ticklabel style={font=\footnotesize},
    height=4.605cm,
    width=8.5cm,
    mark repeat=5,
    mark size=1.5pt,
    mark options={solid}
]
\addplot [very thick, palette-1,mark=diamond*,mark phase=3]
table {%
1 0.0293850898742676
2 3.95774841308594e-05
3 8.0108642578125e-05
4 0.000214576721191406
5 0.000490903854370117
6 0.000936985015869141
7 0.00228643417358398
8 0.00368237495422363
9 0.00728178024291992
10 0.0145654678344727
11 0.0312435626983643
12 0.0570995807647705
13 0.0822162628173828
14 0.136398315429688
15 0.195007562637329
16 0.278795480728149
17 0.394420862197876
18 0.520018100738525
19 0.719104290008545
20 0.878378868103027
21 1.11074066162109
22 1.56954431533813
23 1.87512516975403
24 2.40297865867615
25 3.27652072906494
26 4.17122769355774
27 5.25238680839539
28 6.62653374671936
29 8.22884726524353
30 9.98813438415527
31 11.931519985199
32 14.6313467025757
33 17.4265658855438
34 20.6170241832733
35 24.673095703125
36 28.544792175293
37 32.3940906524658
38 37.6184198856354
39 43.6756076812744
40 50.7424471378326
41 58.5783739089966
42 68.0023345947266
43 77.9502482414246
44 91.6087121963501
45 101.335332155228
};
\addlegendentry{MIMO}
\addplot [very thick, palette-2,mark=triangle*,mark phase=4]
table {%
1 0.0195941925048828
2 6.00814819335938e-05
3 8.94069671630859e-05
4 0.00014948844909668
5 0.000290870666503906
6 0.000438928604125977
7 0.000798225402832031
8 0.00150346755981445
9 0.00314807891845703
10 0.00677633285522461
11 0.0143911838531494
12 0.0301852226257324
13 0.0627238750457764
14 0.130430221557617
15 0.271847248077393
16 0.577846765518188
17 1.17780995368958
18 2.48879861831665
19 5.12791347503662
20 10.4661040306091
21 21.4091160297394
22 45.0180292129517
23 90.120085477829
24 188.230348825455
};
\addlegendentry{ORC brute-force}
\addplot [very thick, palette-3,mark=*,mark phase=2]
table {%
1 0.000776290893554688
2 3.24249267578125e-05
3 3.95774841308594e-05
4 6.69956207275391e-05
5 0.000116586685180664
6 0.000142335891723633
7 0.000185966491699219
8 0.000221014022827148
9 0.000344991683959961
10 0.000524044036865234
11 0.000749349594116211
12 0.0010685920715332
13 0.0013585090637207
14 0.00174665451049805
15 0.00209283828735352
16 0.00270175933837891
17 0.00315308570861816
18 0.00345373153686523
19 0.00415253639221191
20 0.004669189453125
21 0.00502276420593262
22 0.00618410110473633
23 0.00678586959838867
24 0.0077052116394043
25 0.0093228816986084
26 0.0109269618988037
27 0.0121684074401855
28 0.0138819217681885
29 0.0155434608459473
30 0.0177469253540039
31 0.0188701152801514
32 0.021181583404541
33 0.0230135917663574
34 0.0247366428375244
35 0.0274722576141357
36 0.0298495292663574
37 0.0311717987060547
38 0.0334751605987549
39 0.035771369934082
40 0.0357170104980469
41 0.0374643802642822
42 0.0421240329742432
43 0.0473999977111816
44 0.0541133880615234
45 0.0567600727081299
46 0.0593664646148682
47 0.0626394748687744
48 0.0678977966308594
49 0.0736904144287109
50 0.078082799911499
51 0.0846774578094482
52 0.0905113220214844
53 0.0944869518280029
54 0.100986003875732
55 0.10893702507019
56 0.111993789672852
57 0.1213538646698
58 0.123959064483643
59 0.129094362258911
60 0.13700795173645
61 0.143249273300171
62 0.1479332447052
63 0.157366037368774
64 0.162514925003052
65 0.170799493789673
66 0.178555727005005
67 0.189827919006348
68 0.198933362960815
69 0.210114479064941
70 0.22255277633667
71 0.235144376754761
72 0.248618364334106
73 0.255095958709717
74 0.264996767044067
75 0.275715112686157
76 0.288891553878784
77 0.295244693756104
78 0.302764892578125
79 0.312831878662109
};
\addlegendentry{ORC optimized}
\addplot [very thick, palette-4,mark=square*,mark phase=4]
table {%
1 0.190310955047607
2 0.0603291988372803
3 0.0570468902587891
4 0.0512628555297852
5 0.0664160251617432
6 0.0745210647583008
7 0.166727066040039
8 0.195652961730957
9 0.264570474624634
10 0.468659162521362
11 0.714312791824341
12 1.15547966957092
13 1.34865951538086
14 0.383256435394287
15 1.71245718002319
16 2.00445675849915
17 2.1252875328064
18 2.7317488193512
19 2.75267720222473
20 2.69949078559875
21 2.70234107971191
22 2.81834602355957
23 2.84100842475891
24 2.89510011672974
25 2.82813334465027
26 2.85334610939026
27 2.9404981136322
28 2.89021515846252
29 2.91126847267151
30 2.95552563667297
31 2.90436029434204
32 2.89733362197876
33 3.01289558410645
34 3.00857782363892
35 2.93833589553833
36 2.94048190116882
37 3.00819730758667
38 2.95824718475342
39 2.95228171348572
40 2.97833275794983
41 2.94245266914368
42 2.97743725776672
43 2.96329021453857
44 2.98816800117493
45 3.17068767547607
46 3.01167297363281
47 2.99131298065186
48 3.01031589508057
49 3.03349423408508
50 2.99924445152283
51 3.01332521438599
52 3.03631448745728
53 3.10730218887329
54 3.09805798530579
55 3.00743055343628
56 3.1429603099823
57 3.08776664733887
58 3.04831695556641
59 3.05184459686279
60 3.142009973526
61 3.2013053894043
62 3.31426954269409
63 3.3206832408905
64 3.44296550750732
65 3.51593613624573
66 3.56268429756165
67 3.6981987953186
68 3.80442786216736
69 3.85796022415161
70 4.11463260650635
71 4.30703735351562
72 4.38087821006775
73 4.3101692199707
74 4.2942373752594
75 4.35545444488525
76 4.34460759162903
77 4.39376783370972
78 4.3169572353363
79 4.33068370819092
};
\addlegendentry{asclite}

\end{axis}
\end{tikzpicture}

%% file: tikz/runtime_plot_num_speakers_channels.tex
\begin{tikzpicture}

\pgfmathsetmacro\plotw{5cm} 
\pgfmathsetmacro\ploth{4cm} 

\begin{groupplot}[
    group style={
        group size=2 by 2,
        horizontal sep=9pt,
        y descriptions at=edge left,
    },
]
\nextgroupplot[
    xlabel=number of speakers $\nspk$,
    ylabel=runtime in s,
    ylabel style={yshift=-0.5em},
    xmin=1, xmax=9,
    ymin=0.00525, ymax=100,
    ymode=log,
    log basis y={10},
    grid=both,
    grid style={line width=.2pt,draw=black!20},
    legend cell align={left},
    legend style={
      fill opacity=0.8,
      draw opacity=1,
      text opacity=1,
      at={(1.03,1)},
      anchor=south,
      draw=none,
      /tikz/every even column/.append style={column sep=1em}
    },
    legend columns=-1,
    width=\plotw,
    height=\ploth,
    ticklabel style={font=\small},
    mark size=1.5pt,
    mark options={solid}
]
\node[anchor=north west,plotlabel] at (rel axis cs:0,1) {$\nchn=2$};
\addplot[very thick,palette-1,mark=diamond*]
table {%
1 0.0127236127853394
2 0.137404680252075
3 0.853610992431641
4 3.6007194519043
5 14.3020272254944
6 32.4938044548035
7 38.3424162864685
8 171.531762123108
};
\addlegendentry{MIMO}
\addplot [very thick, palette-3, mark=*]
table {%
1 0.0158145332336426
2 0.0109367370605469
3 0.0108823776245117
4 0.0108959674835205
5 0.0114655494689941
6 0.010887622833252
7 0.00985383987426758
8 0.0108847618103027
9 0.00896811485290527
};
\addlegendentry{ORC optimized}
\addplot [very thick, palette-4,mark=square*]
table {%
1 0.249885082244873
2 0.0784752368927002
3 0.197265386581421
4 6.56922459602356
5 39.104035615921
6 246.74041223526
};
\addlegendentry{asclite}
\nextgroupplot[
    xlabel=number of channels $\nchn$,
    xmin=1, xmax=5,
    ymin=0.00525, ymax=100,
    ymode=log,
    log basis y={10},
    grid=both,
    grid style={line width=.2pt,draw=black!20},
    width=\plotw,
    height=\ploth,
    ticklabel style={font=\small},
]
\node[anchor=north west,plotlabel] at (rel axis cs:0,1) {$\nspk=4$};
\addplot [very thick, palette-1, mark=diamond*]
table {%
1 0.0800106525421143
2 3.58418965339661
3 112.517395734787
};
\addplot [very thick, palette-3, mark=*]
table {%
1 0.00612473487854004
2 0.0115163326263428
3 0.344268321990967
4 7.5972888469696
5 152.41418504715
};
\addplot [very thick, palette-4, mark=square*]
table {%
1 2.5350227355957
2 6.39913725852966
3 8.2389087677002
4 16.6573731899261
5 12.9756321907043
6 15.0570282936096
7 10.0396888256073
};
\end{groupplot}

\end{tikzpicture}